\documentclass[preprint,prd,amsmath,nofootinbib,twoside,longbibliography]{revtex4-1}
\usepackage{amsmath}
\usepackage{graphicx}
\usepackage{color}
\usepackage{hyperref}

\begin{document}
\title{New Charged Particles from Higgs Couplings}
\date{July 13, 2012}

\author{Andrew G. Cohen}
\email{cohen@bu.edu}
\author{Martin Schmaltz}
\email{schmaltz@physics.bu.edu}
\affiliation{Physics Department,  Boston University\\
  Boston, MA 02215, USA}

\begin{abstract}
  The recently reported observation of a new particle with mass about
  $125$ GeV and couplings generally resembling those of the Standard
  Model Higgs boson provides a potential probe of the physics of
  electroweak symmetry breaking. Although the current data only
  provides hints, we suggest a particular combination of Higgs
  couplings as an assay for new charged particles connected with
  electroweak symmetry breaking, and construct a simple model with
  charge $5/3$ quarks as a demonstration of its use.
\end{abstract}

\maketitle{}

\section{Introduction and Conclusions}

The CMS and ATLAS collaborations have recently reported\cite{CMS_125,
  *ATLAS_125} the near-discovery of a new particle whose properties
are generally consistent with those of the Standard Model (SM) Higgs
boson. Such a discovery would herald the start of a new era in our
quest to understand the physics of electroweak (EW) symmetry
breaking. While experimenters continue to refine their measurements of
the mass and couplings of this new state, theorists have already begun
the daunting task of interpreting these results in the context of
existing ideas for the origin of electroweak symmetry breaking and
beyond the SM
physics\cite{Carmi:2012in,Espinosa:2012im,Montull:2012ik,Ellis:2012hz,Buckley:2012em,Giardino:2012dp,Corbett:2012dm,Arbey:2012dq,Hardy:2012ef,Baglio:2012et,Gunion:2012gc,Akula:2012kk,An:2012vp,Dev:2012zg}.

What observations indicate that a new particle $h$ is directly
connected with the physics of electroweak symmetry breaking? Broadly
speaking the particle should have interactions reflective of the
electroweak symmetry breaking masses of the SM fermions and gauge
bosons. But somewhat paradoxically some of the best information on the
nature of electroweak symmetry breaking is likely to come from the
couplings to the gauge bosons that do \textit{not\/} acquire
electroweak breaking masses, the gluons and the photon. The couplings
of a Higgs-like state to gluons and photons probe the quantum
corrections coming from the SM fermions and gauge bosons as well as
those from any new states that carry charge or color and couple,
directly or indirectly, to the physics of electroweak symmetry
breaking. Therefore the quantum-induced couplings $h GG$ and $h FF$
are a useful discriminant for models of EW symmetry breaking, can help
test naturalness of this symmetry breaking, and may be one of the
earliest indicators of the existence of charged particles beyond those
of the SM.

Consider the addition to the standard model of a new fermion $\psi$ in
a color representation $R$ carrying electric charge $Q$.  If this new
fermion couples to the physics of electroweak symmetry breaking we
expect its mass to vary with the electroweak-breaking vev $v\sim 246
\text{ GeV}$. But generally not \emph{all\/} of the fermion's mass
will arise from electroweak breaking and the mass parameter is a
function of $v$: $m_{\psi}(v)$. Then the (non-derivative) coupling to
the Higgs\footnote{We simplify the discussion by restricting to a SM
  Higgs doublet. Our discussion is easily generalized to more complex
  Higgs sectors.}  $h$ is $dm_{\psi}/dv$.  Our convention is $m_{\psi}
\ge 0$ so a positive value for $dm_{\psi}/dv$ increases the fermion
mass in the presence of electroweak breaking while a negative value
decreases it.

The contributions of the new fermion $\psi$ to the $hGG$ and $hFF$
couplings at one loop are easily calculated\cite{*[{}] [{ and
  references therein}] Gunion:1989we}.  We
can parameterize the new contributions to $h$ couplings to gluons and
photons as in \cite{Carmi:2012in}:
\begin{equation}
  \label{eq:1}
  \delta {\cal L} = \delta c_{g} \frac{\alpha_{s}}{12 \pi v} h G^{2} + 
  \delta c_{\gamma} \frac{\alpha}{\pi v} h F^{2}
\end{equation}
With this parameterization the top quark loop contributes (in the
limit $2m_{t} \gg m_{h}$)  $\delta c_{g} = 1 \text{
  and } \delta c_{\gamma} = (2/3)^{2}/2 = 2/9$.  For a general fermion
the results are
\begin{equation}
  \label{eq:2}
  \begin{gathered}
    \delta c_{g} = 2\, I(R) \frac{d m_{\psi}}{d v} \frac{v}{m_{\psi}}
    A_{f}(x_{\psi})  = 2\, I(R) \frac{d \ln m_{\psi}}{d \ln v}     
    A_{f}(x_{\psi}) \\
    \delta c_{\gamma} = \frac{Q^{2}}{6} D(R) \frac{dm_{\psi}}{dv}
    \frac{v}{m_{\psi}} A_{f}(x_{\psi}) = \frac{Q^{2}}{6} D(R) \frac{d
      \ln m_{\psi}}{d\ln v} A_{f}(x_{\psi})
  \end{gathered}
\end{equation}
where $I(R) \text{ and } D(R)$ are the index and the dimension of the
color representation $R$ respectively, $x_{\psi} \equiv (m_{h}/2
m_{\psi})^{2}$ and $A_{f}(x) = 3 \left[(x-1)f(x) + x\right]/(2 x^{2})$ where
$f(x) = \arcsin^{2}\sqrt{x}$ for $x \le 1$. Note that $A_{f}(x)$ is very
close to unity for fermion masses larger than the $h$ mass.

The relations eq.~\eqref{eq:2} already provide a useful result: the ratio
of the deviation in the couplings of the $h$ to photons and gluons
is independent of the mass and (non-zero) coupling of the new fermion
and is a measure of the fermion charge:
\begin{equation}
  \label{eq:3}
  \frac{\delta c_{\gamma}}{\delta c_{g}}=  \frac{Q^{2}}{12} \frac{D(R)}{I(R)} 
  = \frac{2}{3} \frac{Q^{2}}{C_2(R)}
\end{equation}
The last equality follows from the relation between the index of a
representation and its quadratic Casimir, $C_2(R) D(R)= I(R) D(G)$
with $D(G)=8$ for $SU(3)$.  A measurement of $\delta c_{\gamma}/\delta
c_{g}$ is a measurement of $Q^2/C_2(R)$ for the heavy fermion, solely
from Higgs physics.

Several comments are in order: 
\begin{enumerate}
\item Increasing the mass of the heavy fermion relative to its EW
  symmetry breaking mass contribution decreases $\delta c_g$ and
  $\delta c_\gamma$ in magnitude. Therefore this method of determining
  the charge of a very heavy fermion requires precise measurements of
  the Higgs couplings.
\item When more than one fermion contributes to eqs.~\eqref{eq:2}
  interpretation of the ratio in eq.~\eqref{eq:3} is not as
  straightforward. In the special case that all heavy fermions whose
  masses receive significant contributions from EW symmetry breaking
  have identical color and electric charges, the charges in
  eq.~\eqref{eq:2} factor out and eq.~\eqref{eq:3} still holds.
\item We may define
  \begin{equation}
    \label{eq:12}
    \overline{q^{2}} \equiv 2 \frac{\delta c_{\gamma}}{\delta c_{g}} \ .
  \end{equation}
  For a single color-triplet fermion this is just the square of the
  electric charge of the fermion. More generally this is an
  ``average'' of the squared charges of multiple fermions weighted by
  $d \ln m/d \ln v$. The quantity $\overline{q^{2}}$ provides a useful
  discriminant among and building guide for models of new physics.
\end{enumerate}

In addition to heavy fermions we may also entertain the possibility of
heavy scalars and vectors whose masses arise in part from EW symmetry
breaking. Similarly to the fermion case, a scalar with mass-squared
$m_S^2(v)$ and a vector with mass-squared $m_V^2(v)$ contribute
\begin{equation}
  \label{eq:4}
  \begin{aligned}
    \delta c_{g} &=\frac{1}{2} I(R_S) \frac{d \ln m_{S}}{d \ln v}
    A_S(x_S)  - \frac{21}{2}  I(R_V) \frac{d \ln m_{V}}{d\ln v}  A_V(x_V) \\
    \delta c_{\gamma} &= \frac{Q_S^{2}}{24} D(R_S) \frac{d \ln
      m_{S}}{d\ln v} A_S(x_S) - \frac{7}{8} Q_V^{2} D(R_V) \frac{d\ln
      m_{V}}{d\ln v} A_V(x_V) \ .
  \end{aligned}
\end{equation}
$R_S$ and $Q_S$ are the color representation and charge of the scalar
and $A_S(x) = 3 \left[f(x) - x\right]/ x^{2}$, and $R_V$, $Q_V$ and
$A_V(x)= \left[3(2x-1)f(x) +3 x+2x^2\right]/(7 x^{2})$ are the
corresponding quantities for the vector.  $A_{S} \text{ and } A_{V}$
both approach one for small arguments. Again for a single particle
(scalar, fermion or vector) the discriminant $\overline{q^{2}}$ is
simply $(4/3) (Q^{2}/ C_{2}(R))$.  In the more general case of several
particles the contribution of scalars to $\delta c$ is a factor of 4
smaller than that of fermions of the same charges, while the
contribution from vectors is more than 5 times larger.  However, most
models of EW symmetry breaking do not contain new colored vectors with
EW symmetry breaking masses so that $I(R_V)=0$, and electrically
charged vectors are strongly constrained from precision EW
measurements, usually requiring masses in the multi-TeV range.

As an example we consider a simple model in which the dominant
contributions to $\delta c_{g}$ and $\delta c_{\gamma}$ stem from a
vector-like colored heavy fermion.  We determine the desired charges
of this new fermion by consulting fits of $\delta c_g$ and $\delta
c_\gamma$ to LHC data. Such fits have been performed by a number of
groups\cite{Carmi:2012in,Espinosa:2012im,Montull:2012ik,Ellis:2012hz,Buckley:2012em}
with generally consistent results.  For example, \cite{Carmi:2012in}
fit $\delta c_g$ and $\delta c_\gamma$ to the combined LHC and
Tevatron cross section data assuming that the tree level couplings of
the Higgs to SM particles maintain their SM values. The fit in Figure
1 of their paper indicates that current data somewhat prefer negative
values for $\delta c_\gamma$ and $\delta c_g$, indicating that the
heavy fermion mass should receive negative contributions from EW
symmetry breaking. Furthermore, the fit shows a correlation between
$\delta c_\gamma$ and $\delta c_g$. Although the current data is not
highly constraining, it does hint at a ratio
\begin{equation}
  \label{eq:5}
  2\frac{\delta c_{\gamma}}{\delta c_{g}} \equiv \overline{q^{2}}
  \agt 2\ .
\end{equation}
For our example of a single fermion dominating the ratio this implies
the electric charge of the fermion should satisfy
\begin{equation}
  \label{eq:6}
  Q^2 \agt \frac32 C_2(R)\ .
\end{equation}

The quadratic Casimir for $SU(3)$ representations is bounded below by
$C_2(R)\ge 4/3$ (with equality for a color triplet), and
eq.~\eqref{eq:6} then forces the colored fermion to have charge $\agt
\sqrt{2}$.  In the interest of keeping the electric charges as small
as possible we choose the color-triplet representation: our new
fermion is a ``quark'' with an exotic charge. The null results from
experimental searches for fractionally charged heavy baryons or mesons
suggest that the charges of any color triplet quarks should be
quantized as
\begin{equation}
  \label{eq:7}
  Q = \frac23+ \text{integer} \ .
\end{equation}
The smallest charge greater than $\sqrt{2}$ consistent with
eq.~(\ref{eq:7}) is $Q = 5/3$. These charge assignments then lead to
the prediction $\overline{q^{2}} = 25/9 \simeq 2.78$.

We now present a minimal model for the couplings of this new quark. We
describe all new fermions by left-handed Weyl spinors. We include a
vector-like pair of $SU(2)$-singlet hypercharge $5/3$ quarks $\psi_S
\text{ and } \overline \psi_S$. Renormalizable couplings of these
fermions to the Higgs doublet require additional quark doublets
$\Psi_D \text{ and } \overline \Psi_D$. The new fields then fill out
$(SU(3), SU(2))_{U(1)}$ representations as
\begin{equation}
  \label{eq:11}
  \psi_S=(3,1)_{5/3}, \quad \overline \psi_S=(\overline 3, 1)_{-5/3}, 
  \quad \Psi_D=(3,2)_{7/6},\quad \overline \Psi_D=(\overline
  3,2)_{-7/6}\ .
\end{equation}
The gauge invariant mass terms and Higgs couplings of these new fields
are then
\begin{equation}
  \label{eq:8}
  -M_D \overline \Psi_D \Psi_D - M_S \overline \psi_S \psi_S 
  - \sqrt{2}\lambda \overline \Psi_D H \psi_S - \sqrt{2}\overline
  \lambda\, \overline   \psi_S H^\dagger \Psi_D + \text{h.c.}\ , 
\end{equation}
Bear in mind that bars on the fields are part of the field names (they
do not correspond to complex conjugation) and the coupling constant
$\overline \lambda$ is unrelated to $\lambda$.

The mass matrix and Higgs couplings follow from the substitution
$H=(v+h,0)/\sqrt{2}$. Notice that the charge $2/3$ quark does not
couple to the Higgs field $h$ and therefore does not contribute to
$\delta c_g$ and $\delta c_\gamma$. There are two Dirac charge
$5/3$ quarks which mix after electroweak symmetry breaking. Since they
have the same color and electric charges eq.~\eqref{eq:3} applies and
we confirm the prediction
\begin{equation}
  \label{eq:9}
  2 \frac{\delta c_{\gamma}}{\delta c_{g}} = 2\,\frac{2}{3}\,
\frac{(5/3)^{2}}{(4/3)^{\phantom{2}}} = \frac{25}{9}\ .
\end{equation}

Having fixed the charge from the discriminant $\overline{q^{2}}$ we
turn to the individual photon and gluon couplings.  The contributions
to $\delta c_{\gamma}$ allow us to further fix the fraction of the
fermion masses arising from EW symmetry breaking.  Only the charge $5/3$
quarks contribute and from eq.~\eqref{eq:2} we see that we need to
compute
\begin{equation}
  \label{eq:13}
  \sum \frac{d\ln m}{d\ln v} = \frac{d \ln \det M^{\dagger}M}{d\ln
    v^{2}}\ .
\end{equation}
The charge $5/3$ mass matrix is 
\begin{equation}
  \label{eq:10}
  M_{5/3} =
  \begin{pmatrix}
    M_D & \lambda v \\
    \overline{\lambda} v & M_S 
  \end{pmatrix}
\end{equation}
with determinant $\det M_{5/3} =
(M_{D}M_{S}-\lambda\overline{\lambda}v^{2}) = M m$ where $M\text{ and
}m$ are the masses of the two charge $5/3$ quarks. Putting these
relations together we obtain (in the heavy mass limit $M,m\gg m_{h}$)
\begin{equation}
  \label{eq:14}
  \delta c_{\gamma} = -\frac{\lambda\overline{\lambda} v^{2}}{M
    m}\frac{25}{9}
\end{equation}
This relation clearly exhibits the decoupling property, vanishing as
we take the vectorial masses $M_{D}, M_{S}$ to infinity while holding
the Yukawa couplings fixed. Also note that in the limit that
$M_{D}\to\infty$ this becomes a classic see-saw, and upon integrating
out the heavy state the system reduces exactly to the single fermion
case we described at the beginning of the introduction.

As a numerical example consider $M_{D} = M_{S} = 900$ GeV and $\lambda
= \overline{\lambda} = 1.2$. Then the charge $2/3$ quark has a mass of
$900$ GeV, while the two charge $5/3$ mass eigenstates have masses
about $m \simeq 600$ GeV and $M \simeq 1.2$ TeV.  The couplings $\delta
c_{\gamma} \simeq -.35, \delta c_{g} \simeq -.25$ are then consistent
with the latest LHC data.

We turn to a rough sketch of the collider phenomenology of the model.
The masses of the three new vector-like quarks, two of charge $5/3$
and one of charge $2/3$, are determined by the free parameters $M_D,
M_S, \lambda, \text{ and }\overline \lambda$. They must lie in the
range of a few 100s of GeV to a few TeV in order to significantly
affect the Higgs branching fraction to photons. Furthermore the
structure of the mass matrices forces the charge $2/3$ quark mass
($900$ GeV in the example point) to lie between the charge $5/3$
masses.

\begin{figure}
  \centering
  \def\svgwidth{3in}
  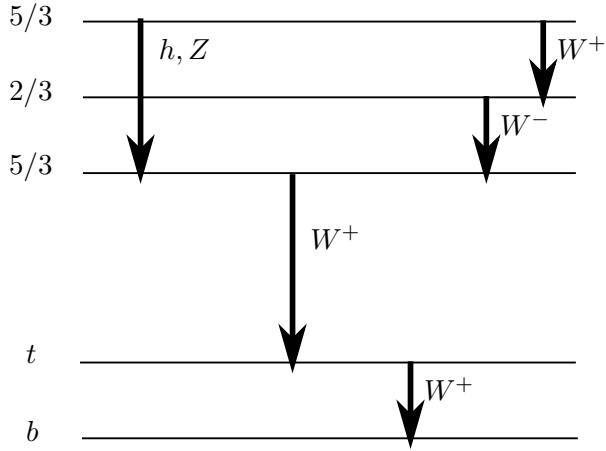
  \caption{Heavy quark mass levels and dominant decays.}
  \label{fig:1}
\end{figure}

At the LHC these new quarks can all be pair-produced with the usual
QCD cross sections. To understand decays note that the only
interactions we have introduced thus far are the gauge interactions
and the Yukawa couplings to the Higgs in eq.~\eqref{eq:8}. These
interactions preserve a heavy quark baryon number under which only the
new quarks are charged. Therefore the lightest charge $5/3$ quark is
stable and the heavier quarks decay to the lightest of the new quarks
by emitting either $W^\pm,Z \text{ or }h$ bosons as shown in
fig.~\ref{fig:1}.  A significant contribution to $\delta c_\gamma$
requires sizeable EW splittings between the heavy quark masses so that
the cascade decays between the heavy quarks produce on-shell $W,Z
\text{ and }h$ particles.

If the lightest new quark were stable, new heavy charged hadrons would
have been observed. Thus we must include interactions which allow the
lightest charge $5/3$ quark to decay into SM particles. There is a
unique renormalizable coupling of the new fermions to the SM fermions:
the ``Yukawa'' coupling of the heavy $SU(2)$ doublet to the SM $SU(2)$
singlet hypercharge $-2/3$ quarks. Given the hierarchical pattern of
Yukawa couplings between the different SM families and given strong
constraints from flavor physics it is natural to expect that the
mixing with first and second family quarks is very small, rendering it
unimportant for LHC phenomenology. So we keep only the coupling to the
top quark
\begin{equation}
  \label{eq:15}
  \lambda_{\text{decay}}  t^c H \Psi_D  \ .
\end{equation}
This coupling ties the heavy quark baryon number to ordinary baryon
number and our heavy triplet quarks then have conventional baryon
number $1/3$.  Upon EW breaking this Yukawa interaction mixes the
heavy charge $2/3$ quark in $\Psi_D$ with the top quark and allows the
heavy quarks to decay to top and bottom quarks through the transitions
indicated in fig.~\ref{fig:1}.

The cascade decays from the heaviest to the lightest of the new quarks
via the gauge couplings and the large Higgs couplings are very
fast. The further decay to SM quarks depends on the arbitrary coupling
$\lambda_{\text{decay}}$: the lightest charge $5/3$ quark may decay
promptly, with a displaced vertex, or it may even be stable on
collider time scales when $\lambda_{\text{decay}} \alt 10^{-8}$.

Different regimes for $\lambda_{\text{decay}}$ lead to varied
phenomenologies. For the smallest couplings heavy quarks are
long-lived leading to a rich system of exotic QCD bound states
containing a heavy quark, including charge 2 mesons and charge 3
baryons. For the largest couplings the top quark mixes strongly with
the exotic charge $2/3$ quark leading to significant modifications of
the top quark couplings to the $W$ and $Z$ bosons. Since $b$ quarks do
not have a corresponding partner to mix with the constraints from
$R_b$ and $A_b$ are weaker.

\begin{figure}
  \centering
  \def\svgwidth{5in}
   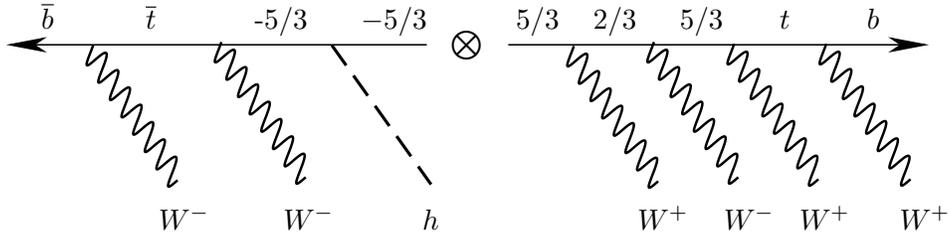
   \caption{Example heavy charge $5/3$ quark decay with six $W$s and a
     Higgs.}
  \label{fig:2}
\end{figure}

We will briefly focus on modest couplings $10^{-6} \alt
\lambda_{\text{decay}} \alt 10^{-1}$ with the lightest new quark decay
being prompt yet still much slower than the decays of the heavier
quarks.  Then the heavier new quarks always decay down to the lightest
new quark, which in turn decays via $W^+ t \rightarrow W^+ W^+
b$. Thus pair production of any of the heavy quarks leads to clear
signatures with 4 to 8 $W$s and 2 $b$ jets, an example of which is
shown in fig.~\ref{fig:2}.

Precision EW constraints on this model are not expected to be very
severe. Contributions of the new fermions to precision observables
arise at loop level and scale like $\delta_{PEW} \sim \lambda^2 /(16
\pi^2) (v^2/M^2)$ which we expect to be sufficiently small for most of
parameter space. Also note that the contributions to the $S$ and $T$
parameters, which usually place the strongest constraints on new
physics coupling to weak gauge bosons, are not enhanced by the square
of the large charges of the new quarks.

The strongest lower bounds on the masses of the heavy quarks come from
ATLAS and CMS searches for fourth family $B$ quarks which are assumed
to decay as $B \rightarrow W^- t \rightarrow W W b$, and therefore $B
\overline{B}$ production leads to the same final state as pair
production of the lighter charge $5/3$ quark. ATLAS searches require
the $B$ mass heavier than about $480$ GeV\cite{ATLAS:2012aw}, while
CMS searches require a mass greater than 611 GeV\cite{Chatrchyan:2012yea}.

\begin{acknowledgments}
  This work was supported by the U.S.\ Department of Energy Office of
  Science.
\end{acknowledgments}

\bibliography{higgs}

\end{document}